\documentclass[aip]{revtex4-1}

\usepackage{epsfig,graphics}
\usepackage{graphicx}% Include figure files
\usepackage{dcolumn}% Align table columns on decimal point
\usepackage{bm}% bold math

\begin{document}

\title[]{Tunable add-drop filter using an active whispering gallery mode microcavity}

\author{Faraz Monifi}
\email{monifi@ese.wustl.edu}
\author{\c{S}ahin Kaya \"{O}zdemir}
\email{ozdemir@ese.wustl.edu}
\author{Lan Yang}
\email{yang@seas.wustl.edu}
\affiliation{Department of Electrical and Systems Engineering, Washington University, St. Louis, Missouri 63130, USA}

%Date{: \today}% It is always \today, today,

\begin{abstract}
An add-drop filter (ADF) fabricated using a whispering gallery mode resonator has different crosstalks for add and drop functions due to non-zero intrinsic losses of the resonator. Here, we show that introducing gain medium in the resonator and optically pumping it below the lasing threshold not only allows loss compensation to achieve similar and lower crosstalks but also tunability in bandwidth and add-drop efficiency. For an active ADF fabricated using an erbium-ytterbium co-doped microsphere, we achieved 24-fold enhancement in the intrinsic quality factor, $3.5$-fold increase in drop efficiency, bandwidth tunability of $35~{\rm MHz}$ and a crosstalk of only $2~\%$.
\end{abstract}

\maketitle
Add-drop systems formed by whispering gallery mode resonators (WGMR)  side-coupled to two waveguides or tapered optical fibers have found applications in a variety of areas, including but not limited to optical sensing, optical communication and signal processing. In particular, there is an increasing interest in their use as optical filters, multiplexers and switches in optical communication networks due to their microscale size, ultra high quality factor $Q$ and large free spectral range.

A WGMR in add-drop configuration- referred to as an add-drop filter (ADF)- selects and transfers a signal light on-resonant with its resonance wavelength either to the add or to the drop port depending on which of the two input ports are used. The signals with off-resonant wavelengths are transferred through the waveguide without coupling into the resonator\cite{1,2,3}. In many realizations of ADFs, the resonance line of the resonator is tuned via thermo-optic effect to provide channel selectivity\cite{4,5}.

An ADF is required to have low crosstalk (ideally zero), and high add-drop efficiencies (ideally unit efficiency). The lower the total loss is, the higher the add-drop efficiency is and also in certain condition the lower the crosstalk is. Therefore, the material of the resonator should be of low-loss in the spectral band of operation, and great care should be taken to minimize the scattering, radiation and coupling losses during the fabrication and coupling processes. Whispering-gallery-mode resonators used in ADFs are typically fabricated from silica, silicon or silicon nitride which have non-zero material loss in the telecommunication band. Moreover, scattering, radiation and coupling losses are fixed at the time of fabrication, so there is not much one can do about these losses after the ADF is fabricated. A mechanism that may help to control or compensate the losses dynamically, even after the fabrication is completed, will be of great interest.

Here we first derive the conditions to achieve zero crosstalk and unit efficiency for a WGMR-based ADF and discuss how one can approach these optimal conditions in practice. We then experimentally demonstrate that a WGMR with tunable optical gain (active WGMR) in ADF configuration helps to dynamically control and compensate a portion of the losses enabling higher add-drop efficiencies and minimal crosstalk, as well as to tune the bandwidth. We also show that with an active WGMRs in ADF, crosstalk values for adding  and dropping can be made the same, which cannot be achieved without the optical gain provided in the active WGMR.

\begin{figure}[]
%\centerline{\includegraphics[width=8cm]{fig1.png}}
\centerline{\includegraphics[width=10cm]{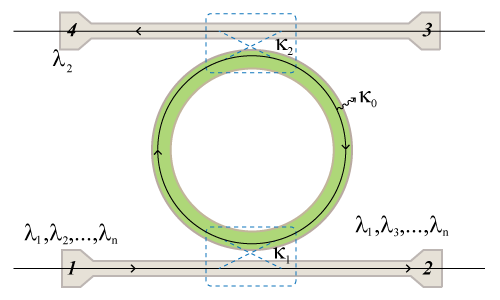}}
\caption{Schematic illustration of an ADF consisting of a WGMR and two waveguides. The waveguide with the ports 1 and 2 is the bus and that with ports 3 and 4 is the drop waveguide. Here the ports 1 and 3 are input ports, and ports 2 and 4 are add and drop ports, respectively. The input field $a_1$ at port 1  consists of a series of signals at wavelengths $ \lambda_1, \lambda_2,...\lambda_n$ whereas the input field $a_3$ at port 3 has $ \lambda_{n+1}, \lambda_{n+2},...\lambda_m$. The signal in port 1 (3) with the wavelength $\lambda_i$ on-resonant with the resonance wavelength of WGMR  couples inside the cavity and is dropped (added) to the field $a_4$ ($a_2$) at port 4 (2). Signals with off-resonant wavelengths are transmitted through the bus and drop waveguides, respectively, to ports 2 and 4. $\kappa_0$ and $\xi$ are the round-trip intrinsic loss (including material, scattering and radiation losses) and gain, respectively. $\kappa_1$ and $\kappa_2$ denote the coupling losses of WGMR-bus and WGMR-drop waveguides, respectively. }\label{fig1}
\end{figure}

A schematics of the ADF is given in Fig. \ref{fig1}. Assuming an intrinsic loss (sum of material, scattering and radiation losses) of $\kappa_0$, we find that for drop functionality the transmission coefficient $T_{1\rightarrow 2}=|a_2/a_1|^2$ is given by\cite{4}:
\begin{equation}\label{N01}
    T_{1\rightarrow 2}=\frac{4\Delta^2+(\kappa_0-\kappa_1+\kappa_2)^2}{4\Delta^2+(\kappa_0+\kappa_1+\kappa_2)^2}.
    \end{equation}
where $\kappa_1$ and $\kappa_2$ denote coupling losses between the WGMR and the waveguides, and $\Delta\equiv\omega-\omega_c$ is the detuning between the laser frequency $\omega$ and the resonance frequency $\omega_c$ of the resonator.

 At resonance ($\Delta=0$), we find that Eq. \ref{N01} becomes
\begin{equation}\label{N03}
    T_{1\rightarrow 2}=\frac{(\kappa_0-\kappa_1+\kappa_2)^2}{(\kappa_0+\kappa_1+\kappa_2)^2}.
\end{equation}
Similarly, we find the drop efficiency $D_{1\rightarrow 4}=|d|^2=|a_4/a_1|^2$ at resonance as
\begin{equation}\label{N04}
    D_{1\rightarrow 4}=\frac{4\kappa_1\kappa_2}{(\kappa_0+\kappa_1+\kappa_2)^2}.
\end{equation}
Here $ T_{1\rightarrow 2}$ corresponds to crosstalk in the system, and should be set to zero for zero crosstalk at $\Delta=0$. It is easy to see that zero-crosstalk is achieved for
    \begin{equation}\label{N05}
     \kappa_1=\kappa_0+\kappa_2
     \end{equation}
which is also the condition for critical coupling for the WGMR-bus waveguide coupling.

Performing the same for the add-functionality with the input at port $3$, we find the transmission  $T_{3\rightarrow 4}=|a_4/a_3|^2$ and the adding efficiency $A_{3\rightarrow 2}=|a_2/a_3|^2$ at resonance as
 \begin{equation}\label{N06}
   T_{3\rightarrow 4}=\frac{(\kappa_0-\kappa_2+\kappa_1)^2}{(\kappa_0+\kappa_1+\kappa_2)^2}
\end{equation}
and
\begin{equation}\label{N07}
   A_{3\rightarrow 2}=\frac{4\kappa_1\kappa_2}{(\kappa_0+\kappa_1+\kappa_2)^2}.
\end{equation} Imposing the zero-crosstalk condition $T_{3\rightarrow 4}=0$ here, we find that
 \begin{equation}\label{N08}
     \kappa_2=\kappa_0+\kappa_1
     \end{equation}
which corresponds to the critical coupling condition for the WGMR-drop waveguide coupling.

Comparing Eqs.\ref{N04} and \ref{N07}, we see that add ($A_{3\rightarrow 2}$) and drop ($D_{1\rightarrow 4}$) efficiencies are always equal to each other but attainable efficiency values are ultimately limited by the intrinsic loss $\kappa_0$, which is decided by the fabrication capabilities and material properties. For a constant $\kappa_0$, drop and add efficiencies are maximized when  $\kappa_1=\kappa_2$, and it increases as $\kappa_1$ and $\kappa_2$ increase. Thus one may argue that by choosing $\kappa_1=\kappa_2\gg\kappa_0$, maximal values for drop-add efficiencies can be achieved. Such an approach suggests increasing the coupling losses significantly. However this will decrease Q (increase the linewidth) of the ADF, thereby limiting the number of channels within the available bandwidth. Moreover, achieving $\kappa_1=\kappa_2\gg\kappa_0$ is challenging and may require tunable resonator-waveguide coupling which will bring additional complexity such as integrating thermal, microelectromechanical systems (MEMs) or nanopositioning systems to fine tune the coupling gap between the resonator and the waveguides. Note that thermal and MEMS can be used for on-chip ADFs fabricated using WGMR and coupled waveguides whereas the nanopositioning system is required for ADFs built using WGMR and coupled fiber tapers.

Equations \ref{N03} and \ref{N06} reveal that for any given setting of $\{\kappa_0,\kappa_1,\kappa_2\}$ crosstalk of add and drop functions are different unless WGMR-bus and WGMR-drop waveguides are simultaneously operated at critical coupling (i.e., Eqs. \ref{N05} and \ref{N08}). This, on the other hand, can be satisfied if and only if $\kappa_0=0$. In other words, to achieve zero crosstalk for both the add and the drop functions of the ADF, intrinsic loss $\kappa_0$ should be zero. Apparently, this is very challenging in practice, because one cannot avoid intrinsic losses completely during the fabrication of the ADF. Therefore, strategies to minimize $\kappa_0$ are needed to achieve higher add-drop efficiencies and lower crosstalks for add and drop. In the following, we show experimentally that this can be achieved by providing optical gain to compensate the intrinsic losses completely or partially.

The WGMR used in our experiments is a silica microsphere ($\sim 90 \mu m$ in diameter) doped with erbium (Er$^{3+}$) and ytterbium (Yb$^{3+}$) ions which provide the necessary optical gain. The microsphere was fabricated by heating the tip of a tapered silica fiber using a ${\rm CO_2}$ laser. Silica microspheres of various sizes can be prepared by controlling the reflow time or the initial size of the tip of the tapered fiber \cite{6}. Using the process proposed by Yang {\it et al.}\cite{7}, we prepared a sol-gel solution consisting of Er$^{3+}$ and Yb$^{3+}$ ions with concentrations of $1.4\times10^{19}$ and  $10^{19}$, respectively. The silica microsphere was dipped into this sol-gel solution twice, each time for about five minutes, to have a coating of Er$^{3+}$-Yb$^{3+}$ co-doped sol-gel silica layer on the microsphere \cite{8}. After each dipping, we reflowed the microsphere using the ${\rm CO_2}$ laser in order to smoothen surface and to have the ions penetrate into the microsphere. In order to fabricate two closely spaced and parallel fiber tapers, we applied the well-known heat-and-pull method simultaneously on two single mode fibers placed on the same fiber holder\cite{4}. The result was two fiber tapers of diameter $2-3~ \mu m$ aligned parallel to each other and separated by around $\sim 150 \mu m$. We placed the microsphere between the fiber tapers, and tuned the gap between the microsphere and the fiber taper waveguides by pushing the waveguides closer to the resonator with the help of a fiber-tip controlled by a $3D$ positioning stage.

\begin{figure}[h!t]
\centerline{\includegraphics[width=10cm]{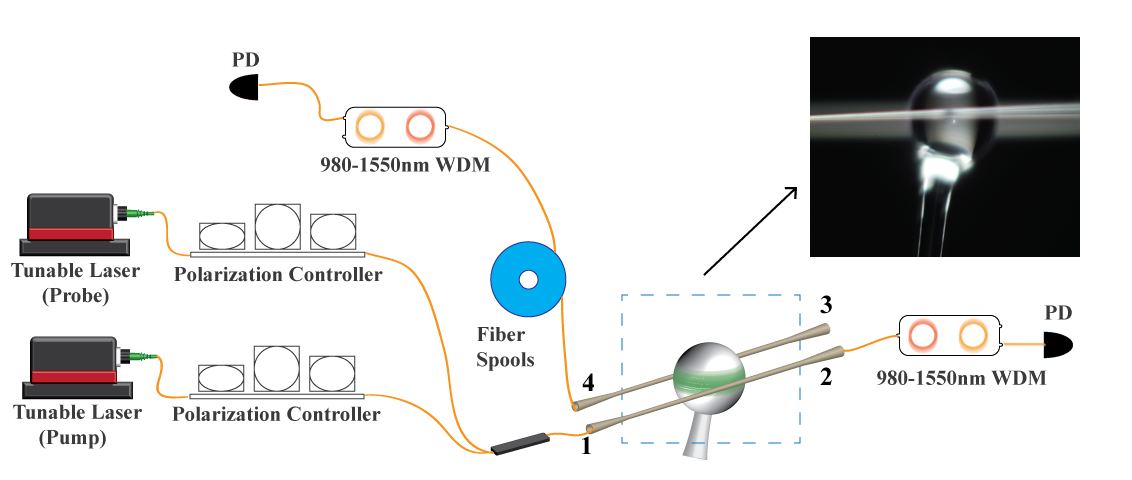}}
\caption{Experimental setup used for the characterization of the add-drop filter (ADF) fabricated from an active Er$^{3+}$-Yb$^{3+}$ co-doped microsphere coupled to two fiber tapers. Probe and pump lasers are in the $1550~ {\rm nm}$ and $980~ {\rm nm}$ bands, respectively. The inset shows an image of the ADF taken \textbf{by} an optical microscope. PD: Photodetector, WDM: Wavelength division multiplexer.}\label{fig2}
\end{figure}

In our system Er$^{3+}$ ions are pumped indirectly via Yb$^{3+}$ which has higher absorption cross-section at $980~ {\rm nm}$, increasing the pump efficiency. Excited Yb$^{3+}$ ions transfer their energies to Er$^{3+}$ ions which then emit light in the $1550~ {\rm nm}$ to compensate the losses in this band\cite{8,9}. The signal (probe) to be added or dropped was in the  $1550~ {\rm nm}$ thereby experiencing the effect of the optical gain. A schematic of our experimental setup is shown in  Fig. \ref{fig2}. The pump laser at $980~{\rm nm}$ and the probe at $1550 {\rm nm}$ were input to the ADF at port 1. We used wavelength division multiplexers to separate the pump and probe lights from each other at ports 2 and 4.

\begin{figure}[]
\centerline{\includegraphics[width=10cm]{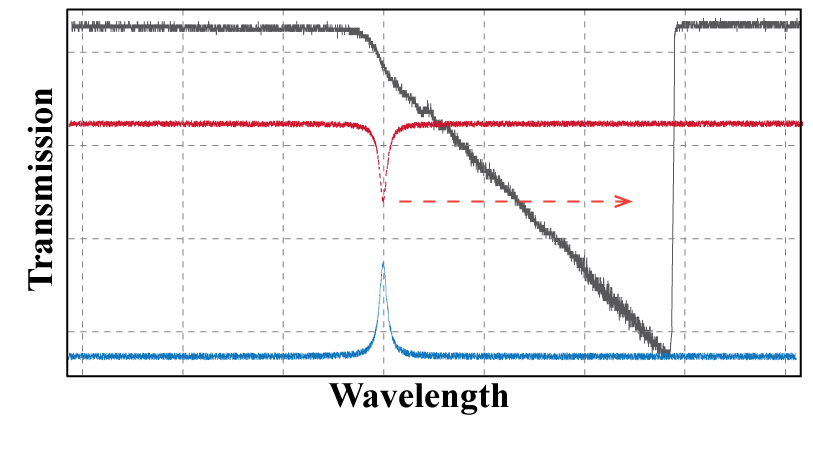}}
\caption{Typical waveforms obtained at ports 2 (throughput) and 4 (drop) when the pump and the weak probe signal are input at port 1 of the add-drop filter and their frequencies are scanned. The high power pump light leads to thermal broadening (black waveform) resulting in the sawtooth-like resonance shape in contrast to a Lorentzian shape obtained for low powers. Along the linear region of the resonance shape, intracavity pump power and hence the optical gain provided by Erbium ions increases linearly. The horizontal dashed orange arrow denotes the direction of increasing pump power. Red and blue colored waveforms with Lorentzian shape correspond to the signals obtained at ports 2 and 4 for the weak probe light at a specific value of the pump power. The $Q$ of the weak probe signal is $6 \times 10^6$}\label{fig3}
\end{figure}

To observe the behavior of the probe in response to changes in the pump power, we scanned the pump laser over $10~{\rm GHz}$ band and focused on a single high-$Q$ mode. For a sufficiently strong pump, thermal broadening kicks in during up-scanning and the Lorentzian resonance mode transforms into a sawtooth like waveform.  Along the slope of this broadened resonance line, the pump power inside the resonator linearly increases as the pump wavelength is scanned (see  Fig. \ref{fig3}). Thus, by scanning the wavelength of the pump laser, we effectively increase the pump power inside the cavity and hence increase the optical gain\cite{10}. In our experiments, we kept the pump power below the lasing threshold. The probe signal power was kept sufficiently small such that it did not introduce thermal nonlinearity and did not affect the thermal broadening of the pump.

\begin{figure}[]
\centerline{\includegraphics[width=10cm]{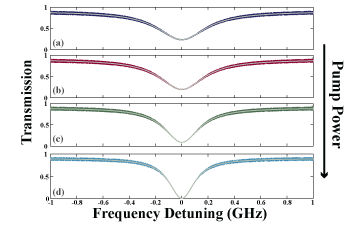}}
\caption{Transmission spectra at the output port 2 when the weak probe was at port 1. Spectra from top to bottom are recorded for increasing pump power. As the power of the pump laser increases, the optical gain provided by the Erbium ions increases compensating a portion of intrinsic losses. This results in $Q$-factor enhancement (linewidth narrowing) and decreased crosstalk. Microsphere-fiber taper coupling shifts from undercoupling (top spectra) to critical coupling (bottom spectra). Dashed white lines represent the best fit to the experimentally obtained spectra.}\label{fig4}
\end{figure}

\begin{figure}[]
\centerline{\includegraphics[width=10cm]{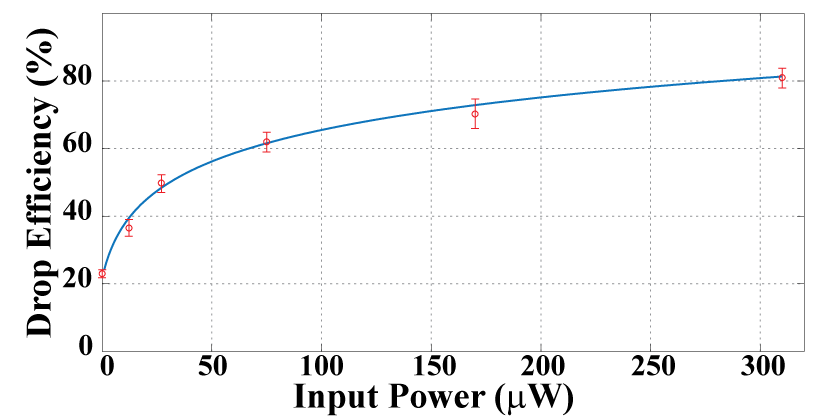}}
\caption{Effect of optical gain provided by Erbium ions on the add-drop efficiency. As the pump power increases, gain compensates losses and shifts the coupling closer to critical coupling and hence decreases crosstalk and increases efficiency. The solid curve denotes the best fit to experimentally obtained data.}\label{fig5}
\end{figure}

Transmission spectra  $T_{1\rightarrow 2}$ obtained at the output port 2 when the input was at port 1 are depicted in  Fig. \ref{fig4} as a function of the pump power. In the following, pump power denotes the power measured at the input of the fiber taper, and is not the intracavity pump power which is much higher than this input power. The WGMR-waveguide coupling was initially set to undercoupling region ($\kappa_1<\kappa_2+\kappa_0$ and $\kappa_2<\kappa_1+\kappa_0$) when the pump was off. The gap between the waveguides and the microsphere were kept constant during the experiment. As the pump power was increased, the spectra showed two significant changes as depicted in \ref{fig4} and \ref{fig5}. First is the gain-induced linewidth narrowing ($Q$-factor enhancement) from $ ~72 MHz$ to $~38.7 MHz$ which is the result of almost $24$ fold decrease in intrinsic loss. Second is the increase in the depth of the resonance dip from $25~\% $ to $ 2~\% $, implying a shift of the coupling condition from undercoupling to critical coupling region. As mentioned above, setting the WGMR-waveguide coupling closer to critical coupling reduces the crosstalk. Thus, these observations imply that the provided optical gain compensates a portion of the intrinsic losses $\kappa_0$, resulting in bandwidth reduction and reduced crosstalk.

Figure \ref{fig5} depicts the experimentally obtained drop efficiency  $D_{1\rightarrow 4}$ as a function of the pump power. We observed an increase from $0.23$ to values greater than $0.81$ in the drop efficiency when the pump power is increased from zero to $310~{\rm \mu W}$. A similar increase was observed for the add efficiency $A_{3\rightarrow 2}$. The relation between the pump power and the drop efficiency can be approximated with the function $a\times P^2(b+c\times P)^{-2}+0.25$ where $a=3.13\times10^{16}$, $b=2\times10^9$ and $c=2.27\times10^8$ are the proportionality constants obtained from curve fitting to experimental data, and $P$ represents the pump power measured at the input of the fiber taper.

\begin{figure}[]
\centerline{\includegraphics[width=10cm]{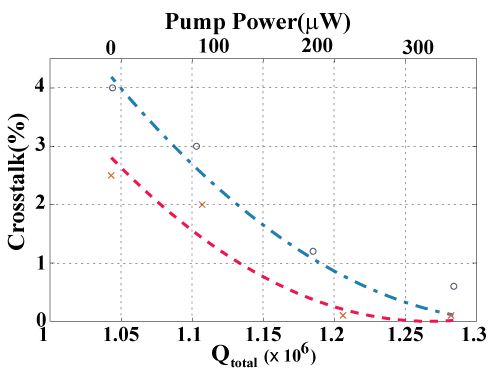}}
\caption{Reduction in crosstalk as a function of pump power (i.e., provided optical gain). As the pump power increases, loaded $Q$ of the system increases and hence the crosstalks for add and drop functions decreases. The data labeled with $\times$ was obtained at port 2 when the weak signal was input at port 1, corresponding to $T_{1\rightarrow 2}$. The data labeled with $o$ was obtained at port 4 when the weak signal was input at port 3, corresponding to $T_{3\rightarrow 4}$. Red dashed and blue dashed-dotted curves are obtained from the theoretical expressions given in Eqs.2 and Eqs.5. }\label{fig6}
\end{figure}

Finally, to show that by compensating the losses, the crosstalks can be balanced and decreased, we put the fiber tapers very close to the resonator ($\kappa_1, \kappa_2 >>\kappa_0$) and increased the pump power gradually. In Fig.  \ref{fig6}, we give the measured crosstalks for the add and drop functionalities as quantified by the depth of the resonance in the transmission spectra $T_{3\rightarrow 4}$ and $T_{1\rightarrow 2}$, respectively. We see that as the pump power increases, loaded $Q$ (i.e., quality factor taking into account all losses including material, scattering, radiation and coupling losses) increases as a result of loss compensation by the optical gain, which in turn, brings our system closer to critical coupling regime and hence leads to deeper resonances in $T_{3\rightarrow 4}$ and $T_{1\rightarrow 2}$.

In conclusion, we have demonstrated that optical gain provided by doped rare-earth-ions in a WGMR used for fabricating add-drop filter can help to increase add and drop efficiency, tune the bandwidth and reduce the cross-talk. Although in this proof-of-principle experiment we have used an Er$^{3+}$-Yb$^{3+}$ co-doped silica microsphere with two coupled fiber taper waveguides, the concept can be extended to other types of WGMRs, such as silica microdisks and silica microtoroids for which gain by rare-earth-ions have already been demonstrated\cite{11,12}, to photonic crystal cavities (PCCs) by doping them with appropriate gain media\cite{13} and to WGMRs and PCCs fabricated from semiconductor materials  such as gallium arsenide, indium gallium arsenide phosphide, etc\cite{14,15}. Similarly, required gain for an active ADF could be provided by quantum dots \cite{16}, other rare-earth-ions\cite{17} or through nonlinear processes, such as Raman gain\cite{18}, and via electrical pumping of cavities fabricated from semiconductor materials.

This work is supported by the US Army Research Office under grant number W911NF-12-1-0026.
%\bibliography{aipsamp}% Produces the bibliography via BibTeX.

\begin{thebibliography}{99}

\bibitem{1} A. S. Kewitsch, G. A. Rakuljic, P. A. Willems, and A. Yariv, \emph{Optics Letters},23, 106, (1998).
\bibitem{2} F. Monifi, A. Ghaffari, M. Djavid, and M. S. Abrishamian, \emph{Appl. Opt.}, 48, 804, (2009).
\bibitem{3} C. Ming, G. Hunziker, and K. Vahala, \emph{IEEE Photon. Technol. Lett.}, 11, 686, (1999).
\bibitem{4} F. Monifi, J. Friedlein, S. K. Ozdemir and L. Yang, \emph{J. Lightw.Technol.}, 30, 21, (2012).
\bibitem{5} B. E. Little et al., \emph{IEEE Photon. Technol. Lett.}, 16, 20, 2263, (2004).
\bibitem{6} M. Cai, O. Painter, K. J. Vahala, and P. C. Sercel, \emph{Optics Letters}, 25, 1430, (2000).
\bibitem{7} L. Yang, D. K. Armani and K. J. Vahala, \emph{Appl. Phys. Lett.}, 83, 825, (2003).
\bibitem{8} L. Yang and K. J. Vahala, \emph{Opt. Lett.}, 24, 592, (2003).
\bibitem{9} H. S. Hsu, C. Cai, and A. M. Armani,  \emph{Opt. Express} 17, 23265 (2009).
\bibitem{10} L. He, S. K. Ozdemir, Y. Xiao and L. Yang, \emph{IEEE J. Quant. Electron.}, 46, 11, (2010).
\bibitem{11} T. J. Kippenberg, J. Kalkman, A. Polman, and K. J. Vahala, \emph{Phys. Rev. A} 74, 051802 (2006).
\bibitem{12} L. He, S. K. Ozdemir and L. Yang, \emph{Laser Photon. Rev.} 7, 60, (2013).
\bibitem{13} M. Makarova, V.Sih, J. Warga, L. Rui, L. Dal Negro and J. Vuckovic, \emph{Appl. Phys. Lett.} , 92, 161107, (2008).
\bibitem{14} U. Mohideen, W. S. Hobson, S. J. Pearton, F. Ren, and R. E. Slusher,  \emph{Appl. Phys. Lett.}, 64, 1911, (1994).
\bibitem{15} M. Loncar, T. Yoshie, A. Scherer, P. Gogna, and Y. Qiu, \emph{Appl.Phys. Lett.} 81, 2680 (2002).
\bibitem{16} S. I. Shopova, G. Farca, A. T. Rosenberger, W. M. S. Wickramanayake, and N. A. Kotov, \emph{Appl. Phys. Lett.} 85, 6101 (2004).
\bibitem{17} J. Wu, S. Jiang, T. Qua, M. Kuwata-Gonokami, and N. Peyghambarian, \emph{Appl. Phys. Lett.} 87, 211118 (2005).
\bibitem{18} S. M. Spillane, T. J. Kippenberg, and K. J. Vahala, \emph{Nature} 415, 621 (2002).
\end{thebibliography}

\end{document}